\documentclass[12pt]{article}

\usepackage[table]{xcolor}

\usepackage{authblk}
\usepackage{soul}

\usepackage{xcolor}
\usepackage{mathpazo, amsmath}
\usepackage{rotating}
\usepackage{url}
\usepackage[sort]{natbib}
\usepackage[T1]{fontenc}
\usepackage[utf8]{inputenc}
\usepackage[left=1in, right=1in, top=1in, bottom=1in]{geometry}
\usepackage{tabularx}

\newcommand{\argmax}{\mathop{\mathrm{argmax}}}

\usepackage{pdflscape}
\usepackage{afterpage}
\usepackage{capt-of}

\usepackage{lipsum}

\usepackage{rotating}

\graphicspath{{./gr/}}

\title{Improvements in Age-Specific Mortality at the Oldest Ages}
\author[1]{Trifon I. Missov}
\author[1]{Silvio C. Patricio}
\author[1,2]{Francisco Villavicencio} 
\affil[1]{\small Interdisciplinary Centre on Population Dynamics, University of Southern Denmark}
\affil[2]{\small Department of Economic, Financial and Actuarial Mathematics, University of Barcelona}
\date{}

\begin{document}
\maketitle
\pagestyle{plain}


\begin{abstract}
Age-specific mortality improvements are non-uniform, neither across ages nor across time. We propose a two-step procedure to estimate the rates of mortality improvement (RMI) in age-specific death rates (ASDR) at ages 85 and above for ten European countries from 1950 to 2019. In the first step, we smooth the raw death counts and estimate ASDR using four different methods: one parametric (gamma-Gompertz-Makeham), two non-parametric ($P$-splines and PCLM), and a novel Bayesian procedure to handle fluctuations resulting from ages with zero death counts. We compare the goodness of fit of the four smoothing methods and calculate the year-to-year ASDR differences according to the best-fitting one. We fit a piecewise linear function to these differences in the second step. The slope in each linear segment captures the average RMI in the respective year range. For each age, we calculate the goodness of fit in the last linear segment to assess how informative the estimated RMI of current mortality change is. The estimated rates of mortality improvement or deterioration (RMI) can be used to make short-term social, health, and social planning, as well as more precise mortality forecasts.
\end{abstract}


\section{\large Rates of Mortality Improvement (RMI)} \label{intro}
Improvements in human survival at older ages result from the extension of lifespans and the postponement of mortality \citep{Zuoetal18}, which is part of a larger life-expectancy revolution \citep{OepVau02}. Deaths are postponed while mortality risks shift toward higher ages, inevitably leading to age-specific mortality improvements at advanced ages \citep{Chretal09, Kanetal94, Rauetal08, Vauetal21}. In most longevous populations today, it also results in an increasing share of nonagenarians and centenarians, whose mortality dynamics influence to a great extent, the changes in the overall death pattern. 

The prospects of longevity, lifesaving, and life expectancy depend on the improvements in age-specific death rates (ASDR), especially at ages above 85, where most deaths in future populations will occur \citep{MesVal00, Vauetal21, Wil00}. Life-table censoring \citep{Misetal16}, scarcity of deaths, and unsatisfactory data quality make estimating mortality progress at these ages complex \citep{Kanetal94, Rauetal08}. In addition there has yet to be a consensus in the literature on the mortality dynamics among the oldest-old. While some studies find evidence for mortality deceleration \citep{HorWil98} even after age 100 \citep{Medetal19}, others point to stagnation in postponing deaths to the oldest ages \citep{Modetal17}. Using Italian data, for instance, Barbi et al. \cite{Baretal18} postulate that the risk of dying closely approaches a plateau after age 105. The statistical model used to arrive at this result, though, has been subjected to criticism \citep{New18a}. Moreover, it has been argued that data errors are the primary cause of the observed late-life mortality deceleration and plateaus \citep{New18b} and that the most recent and reliable data analysis suggests an exponential increase in the risk of death even at very old ages \citep{GavGav19}. Nevertheless, Alvarez et al.\cite{Alvetal21} estimate sex- and age-specific death rates above 105 years using the most recent data from the International Database on Longevity, IDL \cite{IDL21}, with a non-parametric approach: none of the studied populations shows a rapid increase in the hazard of death, and the bigger the sample size for a given country (especially France), the more compelling the evidence of a leveling-off.

The Human Mortality Database, HMD \citep{HMD23}, provides detailed, high-quality harmonized mortality data for a wide range of country-years. However, death rates reported in the HMD result from complex processing of raw data, which is especially significant at older ages where several assumptions are needed \citep{HMDprotocol}. 
To better understand the mortality dynamics at older ages, we apply a two-step procedure to estimate the rates of mortality improvement (RMI) at each age from 85 to 109 in ten European countries. First, we address the problem of data quality in death counts and exposures by applying four approaches to estimate ASDR from raw data. Then, we identify distinct year-ranges of linear increase in ASDR and estimate the slope of each linear segment, which indicates the (average) RMI in the corresponding year-range. The last linear segment reflects the current RMI (CRMI) and the length of the period over which CRMI persists. Depending on the latter, CRMI-estimates can play an essential role in public health strategies and social planning. 


\section{\large Data} \label{data}
This study focuses on Czechia, Denmark, France, Germany, Italy, the Netherlands, Poland, Spain, Sweden, and Great Britain to reflect different types of mortality dynamics, different population sizes, and different sources of data collection (register-based vs. census-based). We use raw death counts and exposures from the Human Mortality Database \citep{HMD23} for years from 1950 to 2019 (for Germany: only 1991 to 2019). We do not include data from 2020 or 2021, where available, as the age-specific death rates in these years are affected by the COVID-19 pandemic. Mortality deterioration due to COVID-19 and its subsequent recovery has been thoroughly studied \citep[see, for example, ][]{Abuetal22, Schetal22}, but it is beyond the scope of this paper. We are interested in the overall trend of RMI, namely whether mortality improvements occur at the oldest ages and how persistent they are. We do not consider data for cohorts because their raw death counts are unavailable, and the resulting death-rate patterns are already smoothed by the HMD \citep{HMDprotocol}. 


\section{\large Methods for Estimating Death Rates from Raw Data} \label{methods}
As reported in the Methods Protocol of the Human Mortality Database, most raw data require various adjustments before being used as inputs to calculate death rates and build life tables. The most common adjustments are distributing deaths of unknown age proportionately across the age range and splitting aggregate data into finer age categories \citep{HMDprotocol}. Among the oldest-old, data quality issues are even more noticeable, with the problem of having zero death counts at some ages. In addition, the HMD makes several assumptions in estimating death rates at older ages. First, observed sex-specific death rates at ages 80 and above are smoothed by fitting a Kannisto model of old-age mortality \citep{Thaetal98}, which is a logistic curve with an asymptote at 1. Fitted rates are used for all ages above 95 years, regardless of the observed death counts. For ages 80--95, within each country-year and sex observed, death rates are used up to the last age $Y$ with at most 100 male or 100 female deaths; observed rates are replaced by the fitted ones for ages above $Y$ \citep{HMDprotocol}.

These and other adjustments in the HMD justify exploring alternative methods to estimate death rates from raw data. Note, for instance, that the Kannisto model implicitly assumes a mortality deceleration at older ages and the existence of a plateau at 1. In the first step, we smooth the raw death counts and estimate ASDR for ages 85--109 using four different methods: one parametric (gamma-Gompertz-Makeham), two non-parametric ($P$-splines and PCLM), and a novel Bayesian procedure to handle fluctuations resulting from ages with zero death counts. We compare the goodness of fit of the four smoothing methods and calculate the year-to-year ASDR differences according to the best-fitting one. In the second step, we fit a piecewise linear function to these differences. We carry out all our analyses using the open-source statistical software R \citep{RCoreTeam}.


\subsection{Gamma-Gompertz-Makeham model}

The gamma-Gompertz-Makeham ($\Gamma$GM) is a parametric mortality model that has been widely used in the literature \citep[see, for instance,][]{VauManSta79, VauMis14}. It is a more flexible version of the Kannisto model \citep{Thaetal98} used by the HMD that allows for any positive asymptote. The mortality hazard of the $\Gamma$GM model is given by

\begin{align*}
	\mu_x = \frac{\alpha\, e^{\beta x}}{1+\frac{\gamma \alpha}{\beta}\left(e^{\beta x}-1\right)}+c\;,
\end{align*}

\noindent where $x\geq 0$ denotes age, and $\alpha, \beta>0$ and $c, \gamma\geq 0$ are parameters. It is based on the Gompertz model with baseline mortality $\alpha$ and rate of aging $\beta$, with the additional feature of capturing the extrinsic mortality (by the Makeham term $c$) and unobserved heterogeneity (frailty), which is assumed to be gamma distributed with unit mean and variance $\gamma$ \citep{VauManSta79}.

The fitting procedure assumes that death counts come from a Poisson distribution with a rate parameter $E_x \mu_x$. Let $D_x$ be the number of deaths in a given age interval $[x, x+1)$ for $x=85,\ldots,109$, and $E_x$ the corresponding exposures. For each country-year and sex, we maximize the Poisson log-likelihood

\begin{align*}
\ln \mathcal{L}(\alpha, \beta, c, \gamma; x)=\sum_{x=85}^{109} \left(D_x\ln \mu_x - E_x \mu_x\right)
\end{align*}

For further discussion on the $\Gamma$GM model and its applications, readers are referred to \cite{VauMis14}, \cite{MisNem15}, and \cite{RibMis16}.


\subsection{Two non-parametric models}

We implement two existing non-parametric models to estimate age-specific death rates from raw death counts and exposures: 1) $P$-splines \citep{EilMar96}, and 2) PCLM, the penalized composite link method \citep{Rizetal15}. Both methods share a common statistical basis, but the latter has been found particularly suitable for reconstructing the tail of a distribution.

\begin{enumerate}
	\item $P$-splines are most frequently used for high-precision smoothing of count data. The method is also based on the assumption that data (in this case, deaths) are Poisson-distributed. We use the R package `MortalitySmooth' \citep{Cam12} to smooth the raw death counts from HMD and estimate the associated ASDR. Readers are referred to \cite{EilMar96} and \cite{Cam12} for additional details.
	\item The PCLM approach is a versatile method to ungroup binned count data, say, age-at-death distributions grouped in age classes. It is based on the idea of $P$-splines and assumes that counts are Poisson-distributed. We use the `ungroup' R package \citep{Pasetal18b} to implement the PCLM. Because of zero deaths at some ages, we first sum up all raw death counts of the oldest age groups. In line with the criterion used by the HMD to estimate death rates \citep{HMDprotocol}, for each country-year and sex, we start the grouping at the first age $Y$ with less than 100 deaths. We then use age-specific death counts for ages 40 to $Y$ from HMD, and the last age-group, $Y+$ with grouped deaths, as an input to PCLM. The PCLM algorithm returns age-specific death counts until the 109-110 age group. We finally use the observed exposures 85--109 from HMD to calculate corresponding age-specific death rates. Readers are referred to \cite{Rizetal15} and \cite{Pasetal18b} for additional details.
\end{enumerate}


\subsection{Bayesian approach}
Let us describe in detail the novel Bayesian approach developed for this paper. Suppose $D_x$ is the number of deaths in a given age interval $[x, x+1)$ for $x = 85, \ldots, 109$. For each $x$, let $D_x$ be Poisson-distributed with $\mathbb{E}D_x = \mathbb{Var}D_x = m_x E_x$, where $m_x$ is the central death rate at age $x$ and $E_x > 0$ denotes exposure in $[x, x+1)$, i.e.,

\begin{align*}
\mathbb{P}(D_x = d) = \frac{(m_x E_x)^d \, e^{-m_x E_x}}{d!} \,.
\end{align*}
For each $x$, the likelihood function of $D_x$ is given by

\begin{align*}
\mathcal{L}(m_x \, | \, D_x = d) = m_x^d \, e^{-m_x E_x} \,.
\end{align*}
Assuming a non-informative or uniform prior distribution for $m_x$, we get a posterior distribution given by

\begin{align*}
f(m_x \, | \, D_x = d) = \frac{E_x^d}{\Gamma(d+1)} \, m_x^{(d+1)-1} \, e^{-m_x E_x} \,,
\end{align*}
which is equivalent to a gamma distribution with parameters $\kappa = d+1$ and $\lambda = E_x$. As a result, to estimate $m_x$, we can use any of the following:

\begin{enumerate}

\item the maximum of the posterior distribution, i.e., $\argmax_{m_x} f(m_x \, | \, D_x = d)$ \\ (equivalent to MLE, the maximum-likelihood estimate) \label{max}

\item the expected value of the posterior distribution, i.e., $\frac{d+1}{E_x}$ \label{mean}

\item the median of the posterior distribution, i.e., $\left\{x : \int\limits_0^x f(m_x \, | \, D_x = d) = 0.5 \right\}$ \label{median}

\end{enumerate}

As we assume that $D_i$ and $D_j$ are independent for any $i\neq j$, we do not impose any structure on the age axis. The likelihood function, from which the posteriori distribution for $m_x$ is built, comes from a single observation. Therefore, despite providing a good approximation for the risk of dying when $D_x=0$, this method might be sensitive to outliers, commonly observed after age 100, given the low corresponding exposures $E_x$.


\section{\large Methods for Estimating Mortality Improvement by Age} \label{methods2}
\begin{figure} 
  \includegraphics[width=\textwidth]{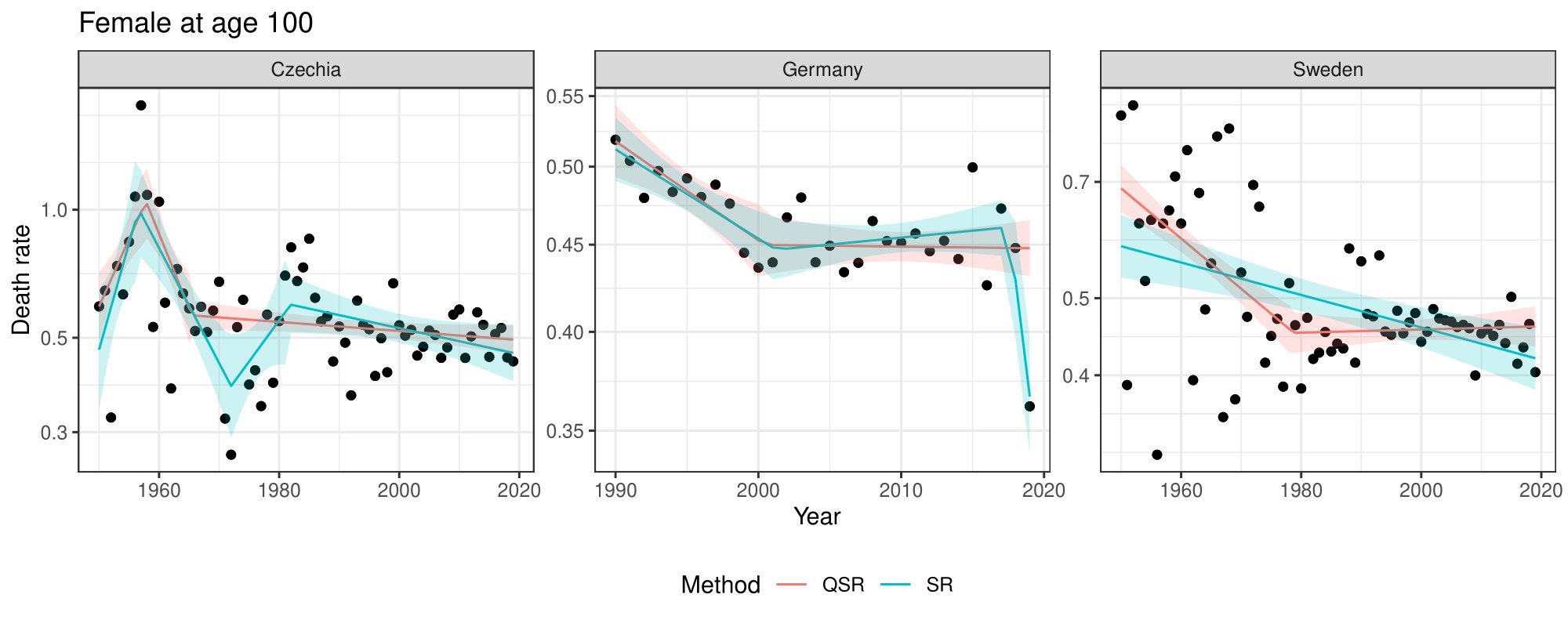}
\caption{Segmented (in blue) vs segmented quantile regression (in red) fits to $\ln m_{100}(t)$-series for females in Czechia, Germany, and Sweden. Examples of the sensitivity of conventional quantile regression to outliers are in the beginning (right panel), in the middle (left panel), and at the end (middle panel) of the study period.}
\label{comp}
\end{figure}

After estimating the death rates at ages 85--109 by the four methods described in Section \ref{methods} (step 1), we use further the $m_x$ estimates according to the best-fitting model. The goodness of fit criterion we apply is the root-mean-square error (RMSE). As year-to-year differences in $m_x$ can fluctuate, even if we take second or higher-order differences, for each $x$, we fit, as a second step, a linear regression to log-mortality for $t = 1950, \ldots, 2019$:

\begin{equation} \label{lr}
\ln m_x(t) = a + bt \,. 
\end{equation}
The slope $b$ accounts for the average rate of mortality improvement (if $b$ is negative) or deterioration (if $b$ is positive). A simple linear model fits well only $\ln m_x$-patterns with a steady trend. When the latter is not present, a natural extension would be to fit a segmented regression \citep{Mug03}, i.e., to assume that $\ln m_x$ has a piecewise-linear structure over time. The slope in the latest time segment would then reflect the average rate of current mortality change (CRMI). Applying a conventional segmented regression might result in too fine partitioning of the year-axis and wide uncertainty intervals for the RMI as the response variable, the expectation of the logarithmic death rates, is sensitive to mortality fluctuations and outliers. As mentioned, the latter is common at the oldest ages with small $E_x$-values. We suggest considering the median (instead of the expectation) of $\ln m_x(t)$ to overcome this problem. The median is still a central tendency measure but also robust to extreme values (outliers). As a result, we fit a linear quantile regression, the median of $\ln m_x(t)$ being the response, with an unknown number of breakpoints. We will call it a \textit{segmented quantile regression}. Even though it has already been applied in \cite{tomal2020ecological}, all statistical properties and technicalities are described in \cite{Patetal23}. Figure \ref{comp} shows how conventional segmented regression responds to outliers at the study period's beginning, middle, and end. When estimating CRMI, a single outlier in the very last year, like the one for German females at age 100 (Figure \ref{comp}, middle panel), creates a new breakpoint in the case of segmented regression. This point defines a new final segment with a steep decline in RMI. On the other hand, the segmented quantile regression remains resistant to this outlier and suggests a much more modest CRMI.

The average rate of mortality improvement in the last log-linear segment, the CRMI, is the most informative regarding mortality forecasts and social and health planning. If researchers and policymakers want to use it, assessing the corresponding goodness of fit is important. For that, we use the metric proposed in \cite{koenker1999goodness}, the pseudo-$R^2$, given by
\begin{equation*}
    R_1(0.5)= 1 - \frac{\sum_{y_i \geq \hat{y}_i}\left|y_i - \hat{y}_i \right|+\sum_{y_i < \hat{y}_i}\left|y_i - \hat{y}_i \right|}{\sum_{y_i \geq \bar{y}_i}\left|y_i - \bar{y}_i \right|+\sum_{y_i < \bar{y}_i}\left|y_i - \bar{y}_i \right|} \,,
\end{equation*}
where $\hat{y}_i$ is the fitted median for the observation $i$, and $\bar{y}_i$ is the fitted value from the intercept-only model. Likelihood ratio tests are carried out using the asymmetric Laplacean density. All technical details can be found in \cite{koenker1999goodness}.


\section{\large RMI by Sex for 10 European Countries in 1950--2019} \label{results}
For each of the ten European countries and in each year from 1950 to 2019, we estimate the age-specific death rates $m_x$ by each of the four smoothing methods described in Section \ref{methods}: $\Gamma$GM, $P$-splines, PCLM and the novel Bayesian procedure. We compare the goodness of fit of the four models by the root-mean-square error (RMSE) and take the smoothed $\ln m_x$ from the best-fitting model (see Table \ref{rmse}). Then, for each series of smoothed $\ln m_x$, we fit a simple linear and segmented quantile regression. We determine the better-fitting regression model by applying a likelihood ratio test (see the resulting piecewise linear fits to the smoothed in step 1 $\ln m_x$ in Figures \ref{f85meanbayes}-\ref{m105meanbayes} of the Appendix). 

\begin{table}[ht] 
\centering
\begin{tabular}{llrrrrl} 
\hline
Sex & Country & $\Gamma$GM & $P$-spline & PCLM & Bayesian & Best \\
\hline
Female & Czechia & 0.71109 & 0.73121 & 0.71189 & 70.58017 & $\Gamma$GM  \\
Female & Germany & 0.15174 & 0.10589 & 0.17938 & 0.08894 & Bayesian \\
Female & Denmark & 0.59029 & 0.60201 & 0.59611 & 36.38742 & $\Gamma$GM  \\
Female & France & 0.50288 & 0.51406 & 0.49271 & 275.39504 & PCLM \\
Female & U.K. & 0.27511 & 0.26141 & 0.28093 & 1.12509 & $P$-spline \\
Female & Italy & 0.38664 & 0.35499 & 0.39471 & 5.74553 & $P$-spline \\
Female & Netherlands & 0.49128 & 0.52711 & 0.50099 & 1456.71440 & $\Gamma$GM  \\
Female & Poland & 0.51929 & 0.52108 & 0.52217 & 55.98287 & $\Gamma$GM  \\
Female & Spain & 0.15198 & 0.10612 & 0.14529 & 0.06501 & Bayesian \\
Female & Sweden & 0.45511 & 0.52355 & 0.46491 & 306.05907 & $\Gamma$GM  \\
Male & Czechia & 0.78001 & 0.79969 & 0.78262 & 578.35638 & $\Gamma$GM  \\
Male & Germany & 0.39312 & 0.35414 & 0.40493 & 5.58635 & $P$-spline \\
Male & Denmark & 0.60277 & 0.72942 & 0.63123 & 85.06176 & $\Gamma$GM  \\
Male & France & 0.52562 & 0.50671 & 0.53359 & 25.46410 & $P$-spline \\
Male & U.K. & 0.39691 & 0.37715 & 0.41085 & 17.78488 & $P$-spline \\
Male & Italy & 0.45973 & 0.43200 & 0.47150 & 11.36622 & $P$-spline \\
Male & Netherlands & 0.58263 & 0.58902 & 0.59399 & 660.47474 & $\Gamma$GM \\
Male & Poland & 0.45450 & 0.44954 & 0.46317 & 65.65627 & $P$-spline \\
Male & Spain & 0.33390 & 0.31520 & 0.37282 & 0.89869 & $P$-spline \\
Male & Sweden & 0.65804 & 0.70141 & 0.67254 & 85.73861 & $\Gamma$GM  \\
\hline
\end{tabular}
\caption{Model-specific root-mean-square errors (RMSE) by country and sex. The best fitting model is listed in the last column.} \label{rmse}
\end{table}

Table \ref{crmi-trust} shows the estimates of CRMI, the average rate of mortality improvement at ages 85, 90, 95, 100, and 105, respectively, by country and sex in the latest time segment (equal to the entire 1950--2019 range if a simple linear regression fits better). Table \ref{crmi-length} contains the lengths of the last linear segment in each case. At age 85, the point estimates vary from $-0.0228$ (Polish females; length of the latest year-segment, $L$, equal to 24 years) to $-0.0108$ (Danish females; $L = 49$). At age 90 mortality progress is more modest: from $-0.0163$ (Polish females; $L = 26$) to $-0.0056$ (Italian females; $L = 15$). At age 95, gains are even smaller: we have CRMI point estimates from $-0.0109$ (Polish females; $L = 25$) to $-0.0028$ (Danish males; $L = 25$). At these ages, the populations in all ten countries, apart from Dutch males at age 90, experience statistically significant mortality improvement (no confidence interval contains 0). CRMI for females slightly dominates the CRMI for males. 

\begin{table}[ht] 
\centering
\resizebox{\linewidth}{!}{%
\begin{tabular}{llccccc}
\hline
Sex & Country & 85 & 90 & 95 & 100 & 105 \\
\hline 
Female & Czechia & \cellcolor{blue!50} -0.0195 (34) & \cellcolor{blue!50} -0.0135 (36) & \cellcolor{blue!40} -0.0092 (35) & \cellcolor{blue!30} -0.0070 (36) & \cellcolor{blue!20} -0.0091 (36) \\
Female & Germany & \cellcolor{blue!40} -0.0137 (20) & \cellcolor{blue!30} -0.0061 (19) & \cellcolor{blue!10} -0.0057 (29) & \cellcolor{blue!10} -0.0046 (29) & 0.0047 (29)  \\
Female & Denmark & \cellcolor{blue!40} -0.0108 (49) & \cellcolor{blue!40} -0.0087 (50) & \cellcolor{blue!50} -0.0076 (25) & -0.0011(48) & 0.0001 (69) \\
Female & France & \cellcolor{blue!50} -0.0218 (44) & \cellcolor{blue!50} -0.0162 (43) & \cellcolor{blue!30} -0.0099 (69) & \cellcolor{blue!10} -0.0071 (69) & \cellcolor{blue!10} -0.0077 (69) \\
Female & U.K. & \cellcolor{blue!50} -0.0139 (41) & \cellcolor{blue!40} -0.0093 (69) & \cellcolor{blue!30} -0.0056 (69) & -0.0004 (34) & \cellcolor{red!10} 0.0021 (37) \\
Female & Italy & \cellcolor{blue!50} -0.0150 (23) & \cellcolor{blue!50} -0.0056 (15) & \cellcolor{blue!50} -0.0064 (25) & \cellcolor{blue!40} -0.0044 (37) & \cellcolor{blue!10} -0.0025 (69) \\
Female & Netherlands & \cellcolor{blue!50} -0.0165 (19) & \cellcolor{blue!50} -0.0092 (21) & \cellcolor{blue!50} -0.0052 (22) & -0.0007 (25) & \cellcolor{red!20} 0.0054 (33) \\
Female & Poland & \cellcolor{blue!50} -0.0228 (24) & \cellcolor{blue!50} -0.0163 (26) & \cellcolor{blue!40} -0.0109 (25) & \cellcolor{blue!10} -0.0014 (61) & 0.0006 (61) \\
Female & Spain & \cellcolor{blue!50} -0.0189 (45) & \cellcolor{blue!50} -0.0127 (49) & \cellcolor{blue!40} -0.0073 (37) & \cellcolor{blue!10} -0.0039 (69) & \cellcolor{red!10} 0.0030 (69) \\
Female & Sweden & \cellcolor{blue!50} -0.0136 (44) & \cellcolor{blue!50} -0.0080 (49) & \cellcolor{blue!40} -0.0038 (45) & 0.0005 (39) & 0.0018 (37) \\
Male & Czechia & \cellcolor{blue!50} -0.0181 (34) & \cellcolor{blue!40} -0.0130 (39) & \cellcolor{blue!40} -0.0073 (29) & -0.0010 (69) & 0.0021 (69) \\
Male & Germany & \cellcolor{blue!40} -0.0129 (14) & \cellcolor{blue!50} -0.0064 (11) & \cellcolor{blue!20} -0.0062 (29) & \cellcolor{blue!10} -0.0026 (29) & 0.0047 (29) \\
Male & Denmark & \cellcolor{blue!50} -0.0194 (17) & \cellcolor{blue!50} -0.0099 (27) & \cellcolor{blue!30} -0.0028 (49) & \cellcolor{blue!10} -0.0026 (44) & 0.0017 (69) \\
Male & France & \cellcolor{blue!50} -0.0209 (22) & \cellcolor{blue!50} -0.0140 (21) & \cellcolor{blue!30} -0.0067 (69) & -0.0037 \cellcolor{blue!10} (69) & 0.0022 (69) \\
Male & U.K & \cellcolor{blue!50} -0.0120 (9) & \cellcolor{blue!50} -0.0144 (21) & \cellcolor{blue!30} -0.0061 (69) & \cellcolor{blue!10} -0.0033 (69) & -0.0004 (69) \\
Male & Italy & \cellcolor{blue!50} -0.0158 (39) & \cellcolor{blue!50} -0.0085 (30) & \cellcolor{blue!50} -0.0042 (32) & \cellcolor{blue!10} -0.0035 (69) & -0.0018 (69) \\
Male & Netherlands & \cellcolor{blue!50} -0.0116 (9) & -0.0069 (10) & \cellcolor{blue!40} -0.0045 (26) & -0.0016 (25) & \cellcolor{blue!10} -0.0017 (69) \\
Male & Poland & \cellcolor{blue!50} -0.0190 (27) & \cellcolor{blue!50} -0.0147 (27) & \cellcolor{blue!40} -0.0071 (20) & \cellcolor{red!10} 0.0028 (51) & 0.0016 (61) \\
Male & Spain & \cellcolor{blue!50} -0.0197 (16) & \cellcolor{blue!40} -0.0101 (46) & \cellcolor{blue!40} -0.0070 (20) & 0.0020 (45) & 0.0052 (40) \\
Male & Sweden & \cellcolor{blue!50} -0.0170 (24) & \cellcolor{blue!50} -0.0106 (18) & \cellcolor{blue!50} -0.0063 (18) & \cellcolor{blue!10} -0.0012 (69) & 0.0010 (69) \\
\hline
\end{tabular}}
\caption{Rates of mortality improvement in the last estimated linear segment (CRMI) by country, sex, and age. Statistically significant mortality improvements are presented in blue, while statistically significant mortality increases are designated in red. The scales of blue and red designate the range of the estimated pseudo-$R^2$: the darkest blue (e.g., Czech females at age 85) designates values $>0.9$, the second darkest blue (e.g., German females at age 85) designates values from $0.8$ to $0.89$, the medium blue scale (e.g., German females at age 90) designates values from $0.7$ to $0.79$, the second lightest blue (e.g., German males at age 95) designates values from $0.6$ to $0.69$, and the lightest blue (e.g., German females at age 95) designates values from $<0.6$. Darker red (Dutch females at age 105 only) designates pseudo-$R^2$ from $0.6$ to $0.69$, while lighter red designates values $<0.6$. The average length (in years) of the last linear segment resulting from fitting a segmented quantile regression to the estimated (by the best-fitting model in Section \ref{methods}) $\ln m_x$, $x = 85, 90, 95, 100, 105$ is presented in brackets.} \label{crmi-trust}
\end{table}

\begin{figure} \centering
 \includegraphics[width=\textwidth]{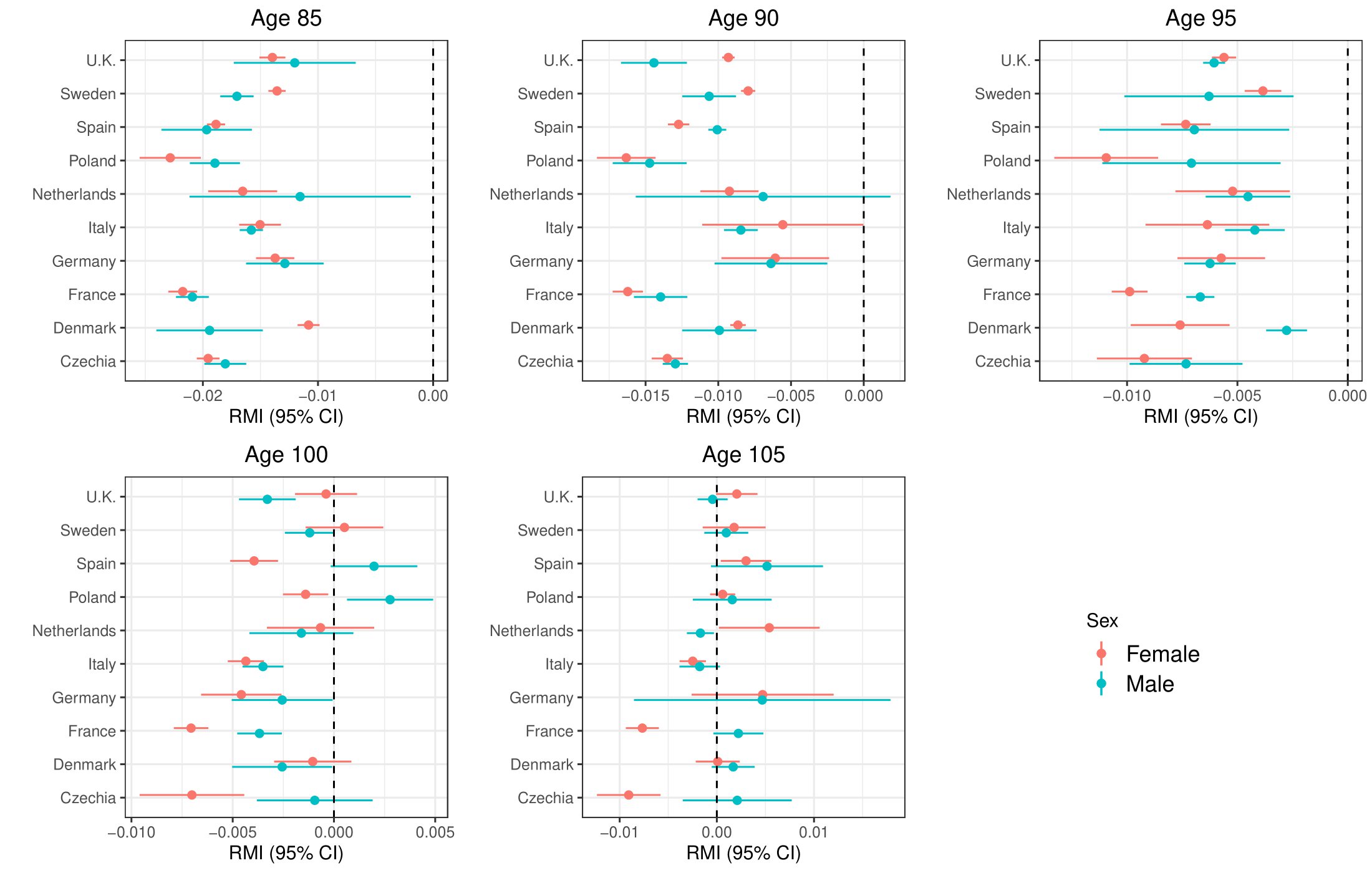}
 \caption{The average rate of mortality improvement at ages 85, 90, 95 (top row), 100 and 105 (bottom row) with 95\% uncertainty intervals (males in blue, females in red), calculated as the slope of a segmented quantile regression for the median of $\ln m_x$, estimated by the best-fitting procedure according to Table \ref{rmse}.}
\label{b85-105}
\end{figure}

At age 100 (Figure \ref{b85-105}, left panel in the second row), 12 out of 20 populations show statistically significant mortality improvement, the CRMI point estimates varying from $-0.0071$ (French females; $L = 69$) to $-0.0012$ (Swedish males; $L = 69$). Seven of the remaining populations experience mortality stagnation at age 100, while death rates for Polish males increase in time at a rate of $0.0016$ ($L = 51$). This confirms the findings by \cite{Modetal17} and \cite{Medetal19} that mortality progress at 100, if any, is very slow. At age 105, though, Czech females experience the highest mortality improvement with a CRMI of $-0.0091$, $L = 36$ (Figure \ref{b85-105}, right panel in the second row). Three other populations are making progress in reducing death rates at 105, at a pace equal to $-0.0077$ (French females; $L = 69$), $-0.0025$ (Italian females; $L = 69$), and $-0.0017$ (Dutch males; $L = 69$). Two populations suffer from statistically significant mortality deterioration in the latest time segment: Spanish females with an average CRMI of $0.0030$ and Dutch females with an average CMRI of $0.0054$. The CMRI for all other populations indicates mortality stagnation at age 105, the point estimates showing, in most cases, slight increases in the death rates. 

The estimated CRMI can be meaningful in mortality forecasting, health, and social planning if a linear model fits the data in the last segment well. Table \ref{pseudoR2} presents the corresponding pseudo-$R^2$ values by country, sex, and age. While the linear model captures with high precision CRMI dynamics at ages 85 and 90 (most pseudo-$R^2$ values being higher than $0.9$, see Table \ref{crmi-trust}), it becomes less accurate at age 95. In contrast, at ages 100 and 105, it fits well only for a handful of populations. This implies that it is safe to use the estimated CRMI only at ages 85, 90, and perhaps 95. In contrast, at ages 100 and 105, researchers and policymakers may consider CRMI-estimates only for those populations where the pseudo-$R^2$ values are high enough (for the purpose CRMI are used). 


\begin{table}[ht] 
\centering
\begin{tabular}{llrrrrr}
\hline
Sex & Country & 85 & 90 & 95 & 100 & 105 \\ 
\hline
Female & Czechia & 0.94446 & 0.91669 & 0.86788 & 0.71137 & 0.63356 \\ 
Female & Germany & 0.84144 & 0.73801 & 0.38764 & 0.29491 & 0.06518 \\ 
Female & Denmark & 0.89872 & 0.89947 & 0.93484 & 0.63266 & 0.00032 \\ 
Female & France & 0.94686 & 0.92820 & 0.78929 & 0.57838 & 0.28778 \\ 
Female & U.K. & 0.93664 & 0.83984 & 0.72746 & 0.78539 & 0.58929 \\ 
Female & Italy & 0.97727 & 0.97716 & 0.94163 & 0.85876 & 0.09755 \\ 
Female & Netherlands & 0.96395 & 0.93636 & 0.90484 & 0.82543 & 0.68588 \\ 
Female & Poland & 0.94298 & 0.92692 & 0.89830 & 0.02367 & 0.01064 \\ 
Female & Spain & 0.95301 & 0.91274 & 0.81653 & 0.26523 & 0.06264 \\ 
Female & Sweden & 0.94418 & 0.93067 & 0.85591 & 0.73487 & 0.74821 \\ 
Male & Czechia & 0.90544 & 0.87914 & 0.85727 & 0.01085 & 0.00387 \\ 
Male & Germany & 0.88607 & 0.93831 & 0.60820 & 0.12348 & 0.01872 \\ 
Male & Denmark & 0.95580 & 0.91461 & 0.72250 & 0.53745 & 0.00913 \\ 
Male & France & 0.97862 & 0.96984 & 0.74455 & 0.26614 & 0.00672 \\ 
Male & U.K. & 0.99089 & 0.95978 & 0.73379 & 0.28027 & 0.00046 \\ 
Male & Italy & 0.95339 & 0.95062 & 0.91282 & 0.28869 & 0.00436 \\ 
Male & Netherlands & 0.97923 & 0.97503 & 0.88639 & 0.84766 & 0.03200 \\ 
Male & Poland & 0.91317 & 0.91323 & 0.84501 & 0.29961 & 0.00329 \\ 
Male & Spain & 0.97621 & 0.87631 & 0.88286 & 0.58427 & 0.89671 \\ 
Male & Sweden & 0.96361 & 0.96310 & 0.92486 & 0.05610 & 0.01627 \\ 
\hline
\end{tabular}
\caption{Values of pseudo-$R^2$, by age, sex and country, for the last linear year-segment resulting from fitting segmented quantile regression to smoothed logarithmic age-specific death rates.} \label{pseudoR2}
\end{table}


\section{\large Discussion}

The rise of human longevity is one of the major achievements of modern societies. As people live longer and life expectancy increases \citep{OepVau02}, more deaths are concentrated at higher ages \citep{Zuoetal18}, and a more profound knowledge of the mortality dynamics among the oldest-old is necessary. Historically, death rates have improved steadily in many countries thanks to advancements in medicine, sanitation, and lifestyle. However, the degree to which mortality continues to improve after a certain age has been debated among demographers and epidemiologists. 

Estimating the rates of mortality improvement (RMI) at the oldest ages has become increasingly important as most deaths, mainly in high-income countries, already take place above age 85. Japanese females, for instance, reached a life expectancy at birth of 87.74 years in 2020 \citep{HMD23}, so RMI above that age can play a crucial role in social, medical, actuarial, and pension planning. Also, in population forecasting, provided that in societies with very low neonatal, infant, and child mortality levels, future life expectancy gains will mainly depend on improvements in mortality at older ages \citep{Vauetal21}.

The scarcity and quality of historical data at these ages call for using models to estimate age-specific death rates (ASDR). This paper explores four approaches: one parametric (gamma-Gompertz-Makeham), two non-parametric ($P$-splines and PCLM), and a novel Bayesian model. We first choose the best-fitting of the four models for each population to smooth the raw death counts. Then, based on the latter and the exposures, we calculate smoothed ASDR and fit a segmented quantile regression. We get the point estimate of the CRMI, the slope in the last linear segment, and a corresponding uncertainty interval. CRMI reflects current mortality improvement (or deterioration) and is essential in short-term planning and forecasting. Using the estimated CRMI, for instance, in a forecast is sensible when the associated linear trend is persistent. To check the latter, we calculate a pseudo-$R^2$ in the last segment. 

The smoothing step identifies the $\Gamma$GM and the $P$-splines as the best-fitting models to the raw death counts. While the novel Bayesian procedure sometimes provides the lowest RMSE, in most cases, its RMSE-values are extremely large. In ages with zero reported deaths $(D_x = 0)$ in which the respective exposures $(E_x)$ are very small, the expected value of the posterior distribution becomes very large. In general, as we impute a uniform random value $u$ for $E_x$, the Bayesian method estimates $1/u$. As $u$ is small and $D_x = 0$, the Bayesian estimates at this age $x$ are very high, which affects the associated RMSE-value.

Our results and estimated CRMI suggest that up to age 100, age-specific death rates decrease in time in all ten countries. After 100, about half of the countries still experience improvements in mortality while it stagnates or slightly deteriorates in the others. Table \ref{crmi-trust} shows that it is sensible to assume a constant yearly rate of mortality change at ages 85, 90, and 95 in the most recent time segment. The associated pseudo-$R^2$ values are above 0.8 almost across all populations. After age 100, there is a stagnation in reducing death rates, with only a few exceptions. At age 105, some populations even experience mortality deterioration. Note that at ages 100 and 105, a linear approximation does not have solid statistical justification. It is to be expected due to the small number of data points the estimation of ASDR at these ages is based on, which is consistent with previous research that quantified significant uncertainty in ASDR above ages 105 and 110 \citep{Alvetal21, VilAbu21}. As a result, the CRMI-estimates at ages 100 and 105 are non-informative for forecasting and social planning. 

Several factors may contribute to the continued improvement in mortality rates among individuals aged 85 and above. First, advancements in medical technology and treatments have allowed for better management of chronic health conditions, such as heart disease and diabetes, which are common among older adults. In addition, lifestyle factors, such as improved nutrition and increased physical activity, may contribute to better health outcomes in later life. Mortality improvements, however, are not uniform across all population subgroups. Improvements among the oldest-old may be impacted by factors such as access to healthcare, social support, and lifestyle. A related question is whether, as life expectancy increases, the extra years of life are being lived in good health. Studies have shown mixed results depending on the age, population, and measure used \citep{Beletal15, Chretal09}, also conditioned by the inherent uncertainty in health estimates compared to mortality data \citep{Viletal21}.


\section*{\large Acknowledgments}	
The research leading to this publication is part of a project that has received funding from the European Research Council (ERC) under the European Union's Horizon 2020 research and innovation program (Grant agreement No. 884328 – Unequal Lifespans). It is also part of a project that has received funding from the ROCKWOOL Foundation through the research project "Challenges to Implementation of Indexation of the Pension Age in Denmark." SCP gratefully acknowledges the support from AXA Research Fund through funding the "AXA Chair in Longevity Research." FV acknowledges funding from the Spanish State Research Agency under the Ram\'on y Cajal program (grant RYC2021-033979-I).


\section*{\large Conflict of Interests}	
None.


\bibliographystyle{unsrtnat}

\bibliography{b-s-lit}

\begin{thebibliography}{39}
\providecommand{\natexlab}[1]{#1}
\providecommand{\url}[1]{\texttt{#1}}
\expandafter\ifx\csname urlstyle\endcsname\relax
  \providecommand{\doi}[1]{doi: #1}\else
  \providecommand{\doi}{doi: \begingroup \urlstyle{rm}\Url}\fi

\bibitem[Zuo et~al.(2018)Zuo, Jiang, Guo, Feldman, and Tuljapurkar]{Zuoetal18}
W.~Zuo, S.~Jiang, Z.~Guo, M.W. Feldman, and S.~Tuljapurkar.
\newblock Advancing front of old-age human survival.
\newblock \emph{Proceedings of the National Academy of Sciences of USA},
  115\penalty0 (44):\penalty0 11209--11214, 2018.
\newblock \doi{10.1073/pnas.1812337115}.

\bibitem[Oeppen and Vaupel(2002)]{OepVau02}
J.~Oeppen and J.W. Vaupel.
\newblock Broken limits to life expectancy.
\newblock \emph{Science}, 296\penalty0 (5570):\penalty0 1029--1031, 2002.
\newblock \doi{10.1126/science.1069675}.

\bibitem[Christensen et~al.(2009)Christensen, Doblhammer, Rau, and
  Vaupel]{Chretal09}
K.~Christensen, G.~Doblhammer, R.~Rau, and J.W. Vaupel.
\newblock Ageing populations: the challenges ahead.
\newblock \emph{The Lancet}, 374\penalty0 (9696):\penalty0 1196--1208, 2009.
\newblock \doi{10.1016/S0140-6736(09)61460-4}.

\bibitem[Kannisto et~al.(1994)Kannisto, Lauritsen, Thatcher, and
  Vaupel]{Kanetal94}
V.~Kannisto, J.~Lauritsen, A.R. Thatcher, and J.W. Vaupel.
\newblock Reductions in mortality at advanced ages: several decades of evidence
  from 27 countries.
\newblock \emph{Population and Development Review}, 20:\penalty0 793--810,
  1994.
\newblock \doi{10.2307/2137662}.

\bibitem[Rau et~al.(2008)Rau, Soroko, Jasilionis, and Vaupel]{Rauetal08}
R.~Rau, E.~Soroko, D.~Jasilionis, and J.W. Vaupel.
\newblock Continued reductions in mortality at advanced ages.
\newblock \emph{Population and Development Review}, 34\penalty0 (4):\penalty0
  747--768, 2008.
\newblock \doi{10.1111/j.1728-4457.2008.00249.x}.

\bibitem[Vaupel et~al.(2021)Vaupel, Villavicencio, and
  Bergeron-Boucher]{Vauetal21}
J.W. Vaupel, F.~Villavicencio, and M.-P. Bergeron-Boucher.
\newblock Demographic perspectives on the rise of longevity.
\newblock \emph{Proceedings of the National Academy of Sciences of USA},
  118\penalty0 (9):\penalty0 e2019536118, 2021.
\newblock \doi{10.1073/pnas.2019536118}.

\bibitem[Mesl{\'e} and Vallin(2000)]{MesVal00}
F.~Mesl{\'e} and J.~Vallin.
\newblock Transition sanitaire: tendances et perspectives.
\newblock \emph{M\'edecine/sciences}, 16:\penalty0 1161--71, 2000.

\bibitem[Wilmoth(2000)]{Wil00}
J.R. Wilmoth.
\newblock Demography of longevity: past, present, and future trends.
\newblock \emph{Experimental Gerontology}, 35\penalty0 (9--10):\penalty0
  1111--1129, 2000.
\newblock \doi{10.1016/S0531-5565(00)00194-7}.

\bibitem[Missov et~al.(2016)Missov, N\'emeth, and Da\'nko]{Misetal16}
T.I. Missov, L.~N\'emeth, and M.J. Da\'nko.
\newblock How much can we trust life tables? {S}ensitivity of mortality
  measures to right-censoring treatment.
\newblock \emph{Palgrave Communications}, 2:\penalty0 15049, 2016.
\newblock \doi{10.1057/palcomms.2015.49}.

\bibitem[Horiuchi and Wilmoth(1998)]{HorWil98}
S.~Horiuchi and J.R. Wilmoth.
\newblock Deceleration in the age pattern of mortality at olderages.
\newblock \emph{Demography}, 35\penalty0 (4):\penalty0 391--412, 1998.
\newblock \doi{10.2307/3004009}.

\bibitem[Medford et~al.(2019)Medford, Christensen, Skytthe, and
  Vaupel]{Medetal19}
A.~Medford, K.~Christensen, A.~Skytthe, and J.W. Vaupel.
\newblock A cohort comparison of lifespan after age 100 in {D}enmark and
  {S}weden: Are only the oldest getting older?
\newblock \emph{Demography}, 56:\penalty0 665--677, 2019.
\newblock \doi{10.1007/s13524-018-0755-7}.

\bibitem[Modig et~al.(2017)Modig, Andersson, Vaupel, Rau, and
  Ahlbom]{Modetal17}
K.~Modig, T.~Andersson, J.W. Vaupel, R.~Rau, and A.~Ahlbom.
\newblock How long do centenarians survive? {L}ife expectancy and maximum
  lifespan.
\newblock \emph{Journal of Internal Medicine}, 282\penalty0 (2):\penalty0
  156--163, 2017.
\newblock \doi{10.1111/joim.12627}.

\bibitem[Barbi et~al.(2018)Barbi, Lagona, Marsili, Vaupel, and
  Wachter]{Baretal18}
E.~Barbi, F.~Lagona, M.~Marsili, J.W. Vaupel, and K.W. Wachter.
\newblock The plateau of human mortality: Demography of longevity pioneers.
\newblock \emph{Science}, 360\penalty0 (6396):\penalty0 1459--1461, 2018.
\newblock \doi{10.1126/science.aat3119}.

\bibitem[Newman(2018{\natexlab{a}})]{New18a}
S.J. Newman.
\newblock Plane inclinations: A critique of hypothesis and model choice in
  barbi et al.
\newblock \emph{PLoS Biology}, 16\penalty0 (12):\penalty0 e3000048,
  2018{\natexlab{a}}.
\newblock \doi{10.1371/journal.pbio.3000048}.

\bibitem[Newman(2018{\natexlab{b}})]{New18b}
S.J. Newman.
\newblock Errors as a primary cause of late-life mortality deceleration and
  plateaus.
\newblock \emph{PLoS Biology}, 16\penalty0 (12):\penalty0 e2006776,
  2018{\natexlab{b}}.
\newblock \doi{10.1371/journal.pbio.2006776}.

\bibitem[Gavrilov and Gavrilova(2019)]{GavGav19}
L.A. Gavrilov and N.S. Gavrilova.
\newblock New trend in old-age mortality: Gompertzialization of mortality
  trajectory.
\newblock \emph{Gerontology}, 65\penalty0 (5):\penalty0 451--457, 2019.
\newblock \doi{10.1159/000500141}.

\bibitem[Alvarez et~al.(2021)Alvarez, Villavicencio, Strozza, and
  Camarda]{Alvetal21}
J.-A. Alvarez, F.~Villavicencio, C.~Strozza, and C.G. Camarda.
\newblock Regularities in human mortality after age 105.
\newblock \emph{PLoS ONE}, 16\penalty0 (7):\penalty0 e0253940, 2021.
\newblock \doi{10.1371/journal.pone.0253940}.

\bibitem[{International Database on Longevity}(2021)]{IDL21}
{International Database on Longevity}.
\newblock French {I}nstitute for {D}emographic {S}tudies ({INED}) (host), 2021.
\newblock Avialable at \url{https://www.supercentenarians.org/}.

\bibitem[HMD(2023)]{HMD23}
HMD.
\newblock {Human Mortality Database}, 2023.
\newblock Max Planck Institute for Demographic Research (Germany), University
  of California, Berkeley (USA), and French Institute for Demographic Studies
  (France). Available at \url{http://www.mortality.org} (data downloaded on
  February 14, 2023).

\bibitem[Wilmoth et~al.(2021)Wilmoth, Andreev, Jdanov, Glei, Riffe,
  et~al.]{HMDprotocol}
J.R. Wilmoth, K.~Andreev, D.~Jdanov, D.A. Glei, T.~Riffe, et~al.
\newblock Methods protocol for the {H}uman {M}ortality {D}atabase, {V}ersion 6.
\newblock Technical report, University of California, Berkeley, and Max Planck
  Institute for Demographic Research, Rostock, 2021.
\newblock Available at \url{
  https://www.mortality.org/File/GetDocument/Public/Docs/MethodsProtocolV6.pdf}
  (retrieved on March 10, 2023).

\bibitem[Aburto et~al.(2022)Aburto, Sch{\"o}ley, Kashnitsky, Zhang, Rahal,
  Missov, Mills, Dowd, and Kashyap]{Abuetal22}
J.M. Aburto, J.~Sch{\"o}ley, I.~Kashnitsky, L.~Zhang, C.~Rahal, T.I. Missov,
  M.C. Mills, J.B. Dowd, and R.~Kashyap.
\newblock Quantifying impacts of the {COVID-19} pandemic through
  life-expectancy losses: a population-level study of 29 countries.
\newblock \emph{International Journal of Epidemiology}, 51\penalty0
  (1):\penalty0 63--74, 2022.
\newblock \doi{10.1093/ije/dyab207}.

\bibitem[Sch{\"o}ley et~al.(2022)Sch{\"o}ley, Aburto, Kashnitsky, Kniffka,
  Zhang, Jaadla, Dowd, and Kashyap]{Schetal22}
J.~Sch{\"o}ley, J.M. Aburto, I.~Kashnitsky, M.S. Kniffka, L.~Zhang, H.~Jaadla,
  J.B. Dowd, and R.~Kashyap.
\newblock Life expectancy changes since {COVID-19}.
\newblock \emph{Nature Human Behaviour}, 6\penalty0 (12):\penalty0 1649--1659,
  2022.
\newblock \doi{10.1038/s41562-022-01450-3}.

\bibitem[Thatcher et~al.(1998)Thatcher, Kannisto, and Vaupel]{Thaetal98}
A.R. Thatcher, V.~Kannisto, and J.W. Vaupel.
\newblock \emph{The force of mortality at ages 80 to 120}.
\newblock Odense University Press, Odense, Denmark, 1998.

\bibitem[{R Core Team}(2021)]{RCoreTeam}
{R Core Team}.
\newblock \emph{R: A Language and Environment for Statistical Computing}.
\newblock R Foundation for Statistical Computing, Vienna, Austria, 2021.
\newblock URL: https://www.R-project.org/.

\bibitem[Vaupel et~al.(1979)Vaupel, Manton, and Stallard]{VauManSta79}
J.W. Vaupel, K.G. Manton, and E.~Stallard.
\newblock The impact of heterogeneity in individual frailty on the dynamics of
  mortality.
\newblock \emph{Demography}, 16:\penalty0 439--454, 1979.
\newblock \doi{10.2307/2061224}.

\bibitem[Vaupel and Missov(2014)]{VauMis14}
J.W. Vaupel and T.I. Missov.
\newblock Unobserved population heterogeneity: A review of formal
  relationships.
\newblock \emph{Demographic Research}, 31\penalty0 (22):\penalty0 659--686,
  2014.
\newblock \doi{10.4054/DemRes.2014.31.22}.

\bibitem[Missov and N\'emeth(2015)]{MisNem15}
T.I. Missov and L.~N\'emeth.
\newblock Sensitivity of model-based human mortality measures to exclusion of
  the {M}akeham or the frailty parameter.
\newblock \emph{Genus}, 71\penalty0 (2-3):\penalty0 113--135, 2015.

\bibitem[Ribeiro and Missov(2016)]{RibMis16}
F.~Ribeiro and T.I. Missov.
\newblock Revisiting mortality deceleration patterns in a
  gamma-{G}ompertz-{M}akeham framework.
\newblock In Robert Schoen, editor, \emph{Dynamic Demographic Analysis}, pages
  117--146. Springer, Cham, 2016.

\bibitem[Eilers and Marx(1996)]{EilMar96}
P.H.C. Eilers and B.D. Marx.
\newblock Flexible smoothing with {$B$}-splines and penalties.
\newblock \emph{Statistical Science}, 112:\penalty0 89--121, 1996.
\newblock \doi{10.1214/ss/1038425655}.

\bibitem[Rizzi et~al.(2015)Rizzi, Gampe, and Eilers]{Rizetal15}
S.~Rizzi, J.~Gampe, and P.H.C. Eilers.
\newblock Efficient estimation of smooth distributions from coarsely grouped
  data.
\newblock \emph{American Journal of Epidemiology}, 182\penalty0 (2):\penalty0
  138--147, 2015.
\newblock \doi{10.1093/aje/kwv020}.

\bibitem[Camarda(2012)]{Cam12}
C.G. Camarda.
\newblock {MortalitySmooth}: An {R} package for smoothing {P}oisson counts with
  {P}-splines.
\newblock \emph{Journal of Statistical Software}, 50\penalty0 (1):\penalty0
  1--24, 2012.
\newblock \doi{10.18637/jss.v050.i01}.

\bibitem[Pascariu et~al.(2018)Pascariu, Da\'nko, Sch{\"o}ley, and
  Rizzi]{Pasetal18b}
M.D. Pascariu, M.J. Da\'nko, J.~Sch{\"o}ley, and S.~Rizzi.
\newblock ungroup: An {R} package for efficient estimation of smooth
  distributions from coarsely binned data.
\newblock \emph{Journal of Open Source Software}, 3\penalty0 (29):\penalty0
  937, 2018.
\newblock \doi{10.21105/joss.00937}.

\bibitem[Muggeo(2003)]{Mug03}
M.R. Muggeo.
\newblock Estimating regression models with unknown break-points.
\newblock \emph{Statistics in Medicine}, 22\penalty0 (19):\penalty0 3055--3071,
  2003.
\newblock \doi{10.1002/sim.1545}.

\bibitem[Tomal and Ciborowski(2020)]{tomal2020ecological}
J.H. Tomal and J.J.H. Ciborowski.
\newblock Ecological models for estimating breakpoints and prediction
  intervals.
\newblock \emph{Ecology and Evolution}, 10\penalty0 (23):\penalty0
  13500--13517, 2020.
\newblock \doi{10.1002/ece3.6955}.

\bibitem[Patricio et~al.(2023)Patricio, Sarnaglia, and Missov]{Patetal23}
S.C. Patricio, A.J. Sarnaglia, and T.I. Missov.
\newblock Segmented quantile regression.
\newblock \emph{Manuscript in preparation}, 2023.

\bibitem[Koenker and Machado(1999)]{koenker1999goodness}
R.~Koenker and J.A.F. Machado.
\newblock Goodness of fit and related inference processes for quantile
  regression.
\newblock \emph{Journal of the American Statistical Association}, 94\penalty0
  (448):\penalty0 1296--1310, 1999.
\newblock \doi{10.1080/01621459.1999.10473882}.

\bibitem[Villavicencio and Aburto(2021)]{VilAbu21}
F.~Villavicencio and J.M. Aburto.
\newblock Does the risk of death continue to rise among supercentenarians?
\newblock In Heiner Maier, Bernard Jeune, and James~W. Vaupel, editors,
  \emph{Exceptional lifespans}, chapter~4, pages 37--48. Springer, Cham, 2021.
\newblock \doi{10.1007/978-3-030-49970-9_4)}.

\bibitem[Beltr{\'a}n-S{\'a}nchez et~al.(2015)Beltr{\'a}n-S{\'a}nchez, Soneji,
  and Crimmins]{Beletal15}
H.~Beltr{\'a}n-S{\'a}nchez, S.~Soneji, and E.M. Crimmins.
\newblock Past, present, and future of healthy life expectancy.
\newblock \emph{Cold Spring Harbor Perspectives in Medicine}, 5\penalty0
  (11):\penalty0 a025957, 2015.
\newblock \doi{10.1101/cshperspect.a025957}.

\bibitem[Villavicencio et~al.(2021)Villavicencio, Bergeron-Boucher, and
  Vaupel]{Viletal21}
F.~Villavicencio, M.-P. Bergeron-Boucher, and J.W. Vaupel.
\newblock Reply to {P}ermanyer et al.: The uncertainty surrounding healthy life
  expectancy indicators.
\newblock \emph{Proceedings of the National Academy of Sciences of USA},
  118\penalty0 (46):\penalty0 e2115544118, 2021.
\newblock \doi{10.1073/pnas.2115544118}.

\end{thebibliography}

\clearpage

\appendix

\section*{\large Appendix: Estimated $\ln m_x$ and segmented quantile regression fits}

\begin{sidewaysfigure}
  \includegraphics[width=0.9\textwidth]{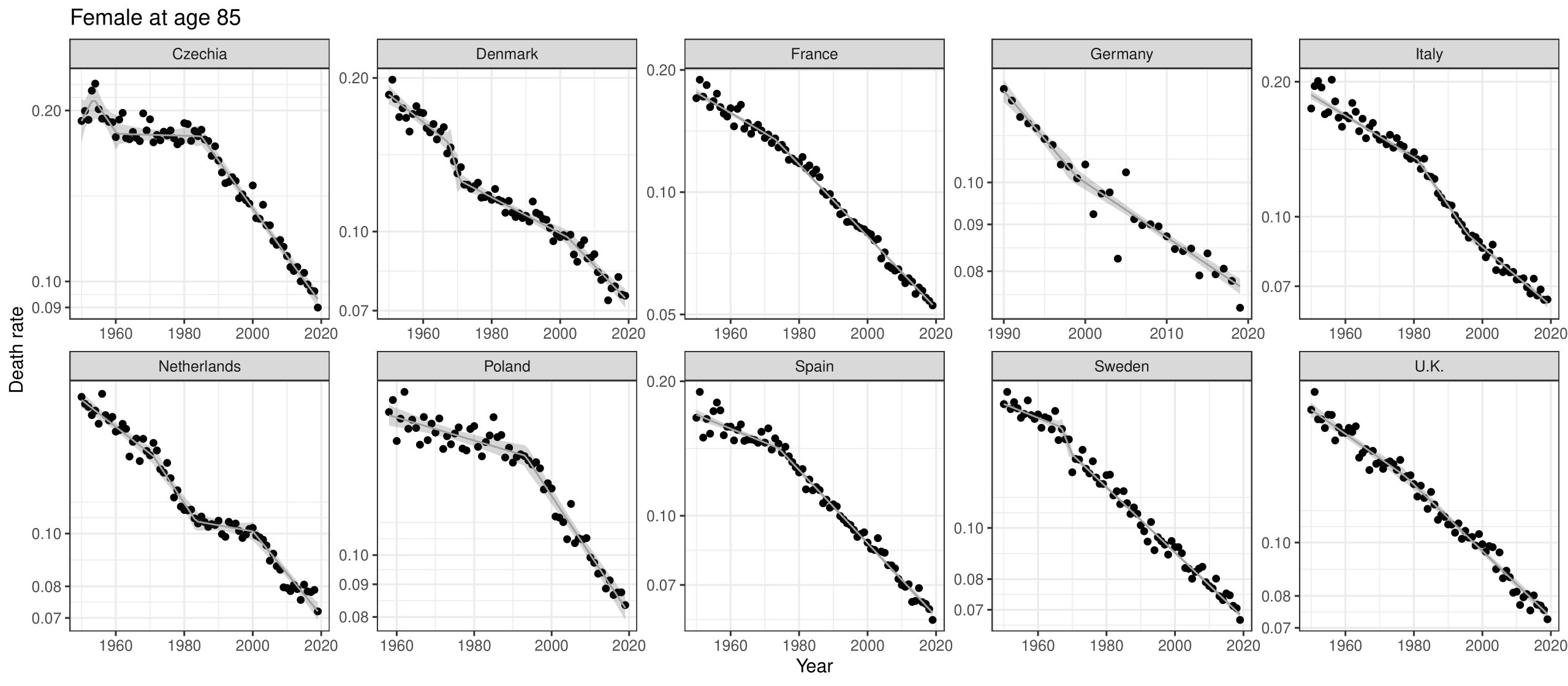}
  \caption{The 1950--2019 (for Germany, 1990--2019) time series of female death rates at age 85 (points), obtained by smoothing the death counts at ages 85 and above by the best-fitting procedure (see Table \ref{rmse}) described in Section \ref{methods}. The solid line represents the best-fitting linear or piecewise linear approximation with its corresponding uncertainty bounds. The existence and number of breakpoints were determined by fitting a segmented quantile regression and testing it (via a likelihood-ratio test) against a linear regression.} \label{f85meanbayes}
\end{sidewaysfigure}

\begin{sidewaysfigure}
  \includegraphics[width=\textwidth]{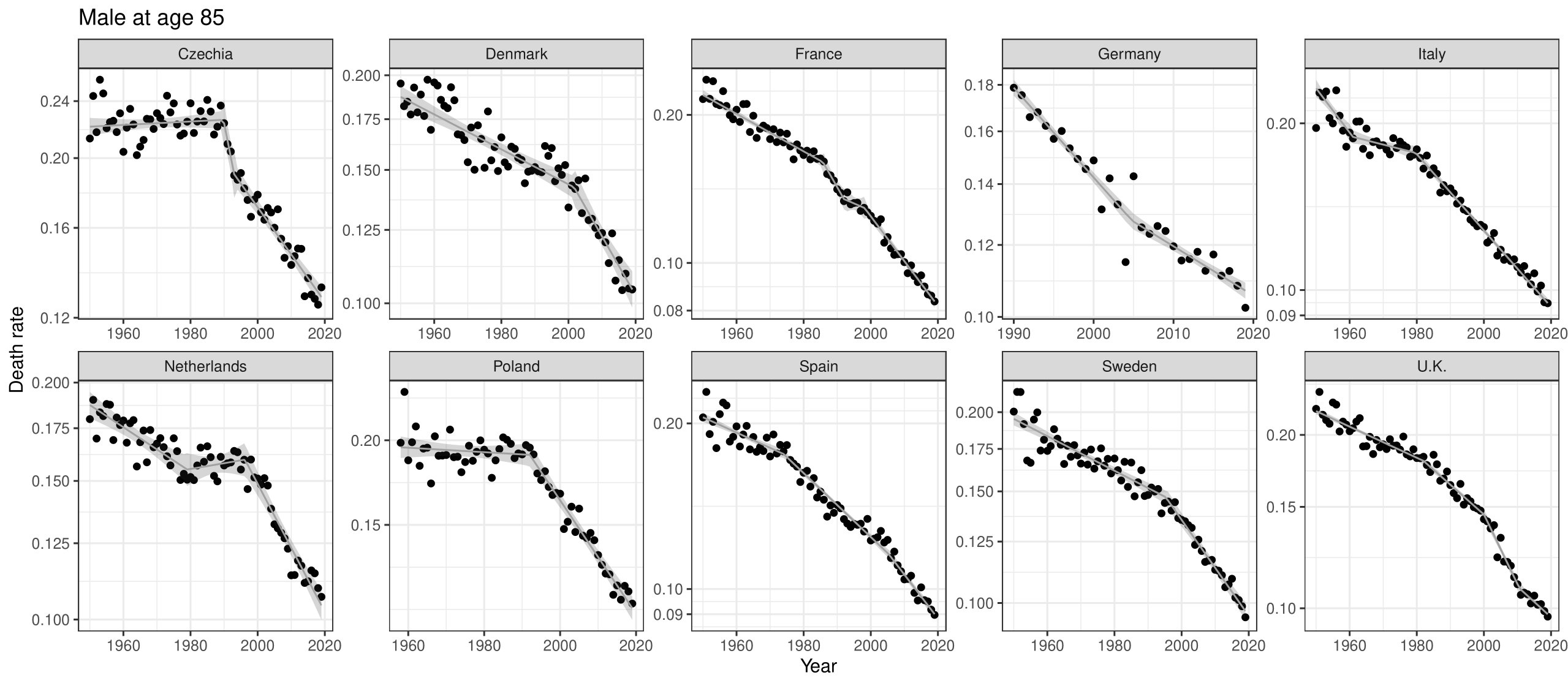}
  \caption{The 1950--2019 (for Germany, 1990--2019) time series of male death rates at age 85 (points), obtained by smoothing the death counts at ages 85 and above by the best-fitting procedure (see Table \ref{rmse}) described in Section \ref{methods}. The solid line represents the best-fitting linear or piecewise linear approximation with its corresponding uncertainty bounds. The existence and number of breakpoints were determined by fitting a segmented quantile regression and testing it (via a likelihood-ratio test) against a linear regression.} \label{m85meanbayes}
\end{sidewaysfigure}

\begin{sidewaysfigure}
  \includegraphics[width=\textwidth]{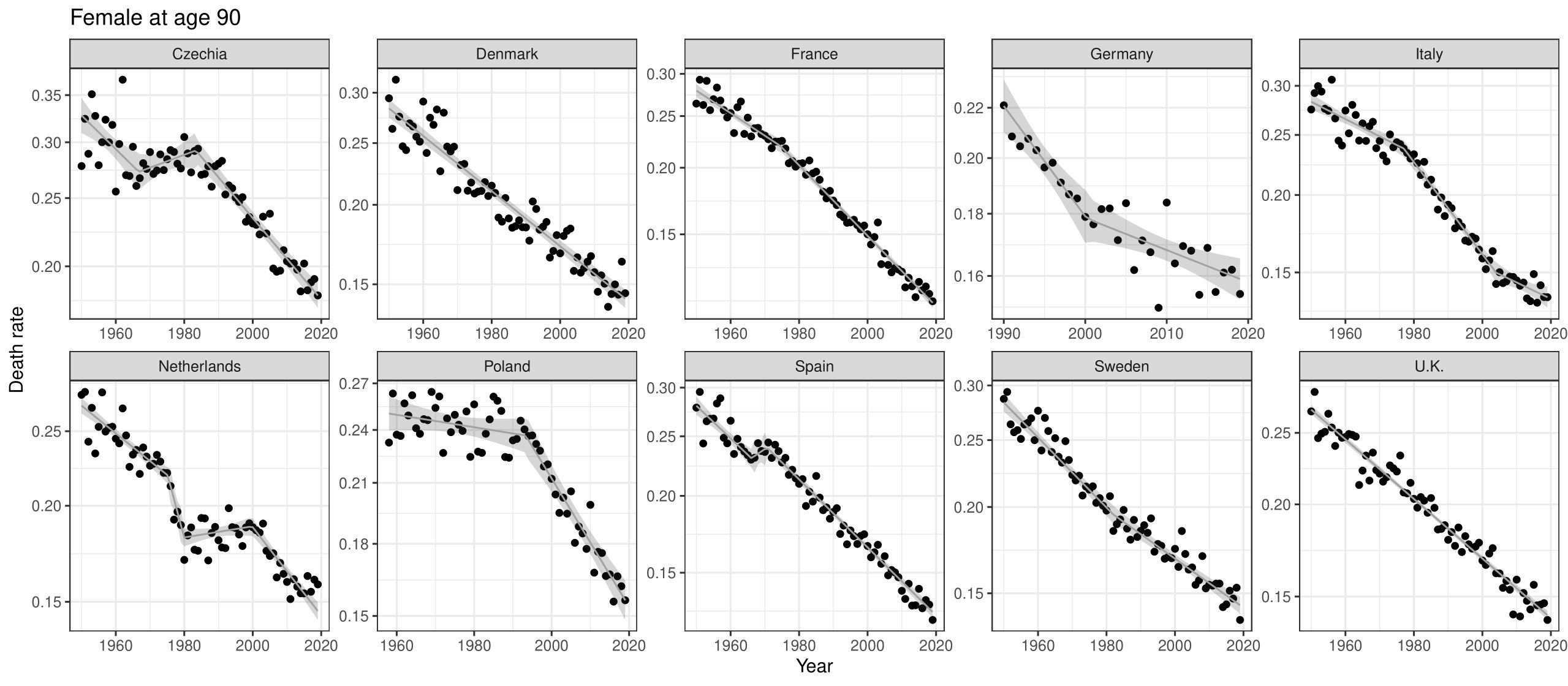}
  \caption{The 1950--2019 (for Germany, 1990--2019) time series of female death rates at age 90 (points), obtained by smoothing the death counts at ages 85 and above by the best-fitting procedure (see Table \ref{rmse}) described in Section \ref{methods}. The solid line represents the best-fitting linear or piecewise linear approximation with its corresponding uncertainty bounds. The existence and number of breakpoints were determined by fitting a segmented quantile regression and testing it (via a likelihood-ratio test) against a linear regression.} \label{f90meanbayes}
\end{sidewaysfigure}

\begin{sidewaysfigure}
  \includegraphics[width=\textwidth]{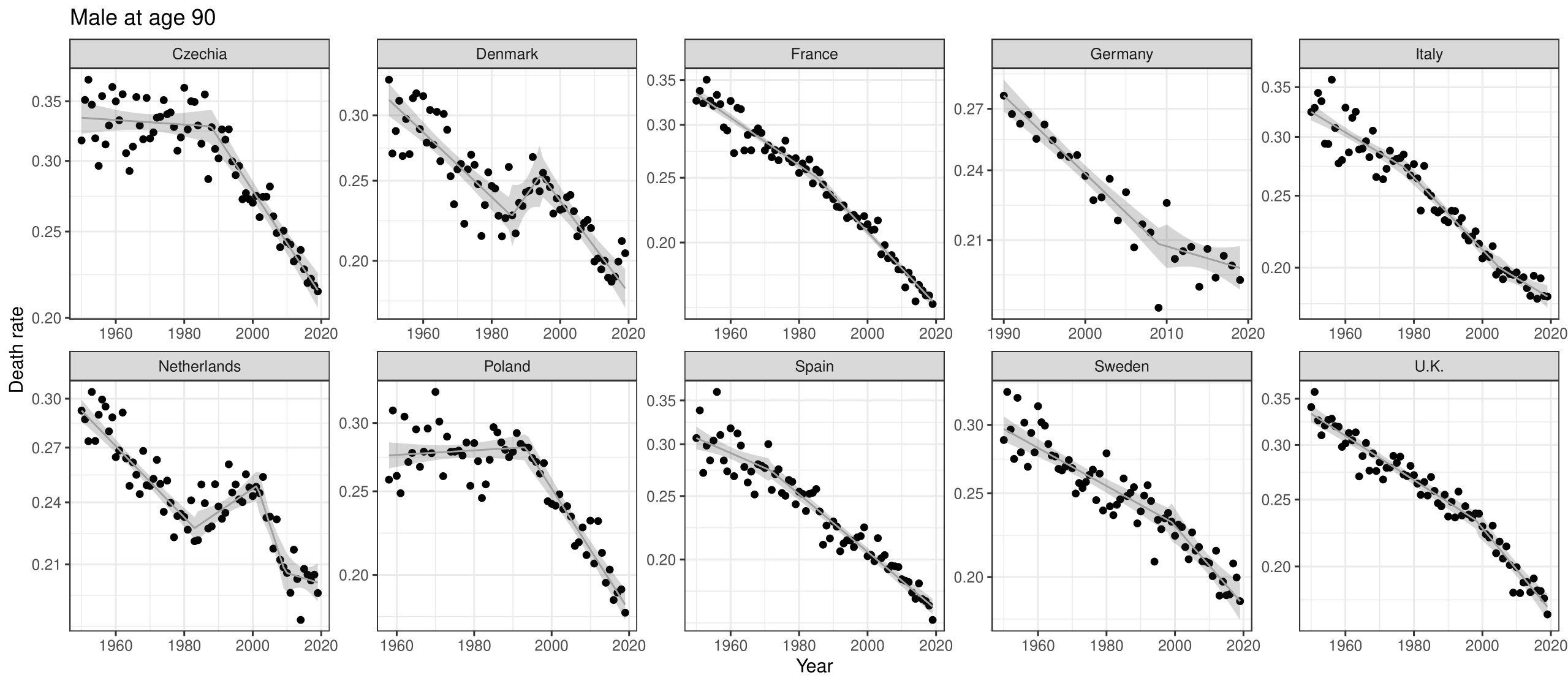}
  \caption{The 1950--2019 (for Germany, 1990--2019) time series of male death rates at age 90 (points), obtained by smoothing the death counts at ages 85 and above by the best-fitting procedure (see Table \ref{rmse}) described in Section \ref{methods}. The solid line represents the best-fitting linear or piecewise linear approximation with its corresponding uncertainty bounds. The existence and number of breakpoints were determined by fitting a segmented quantile regression and testing it (via a likelihood-ratio test) against a linear regression.} \label{m90meanbayes}
\end{sidewaysfigure}

\begin{sidewaysfigure}
  \includegraphics[width=\textwidth]{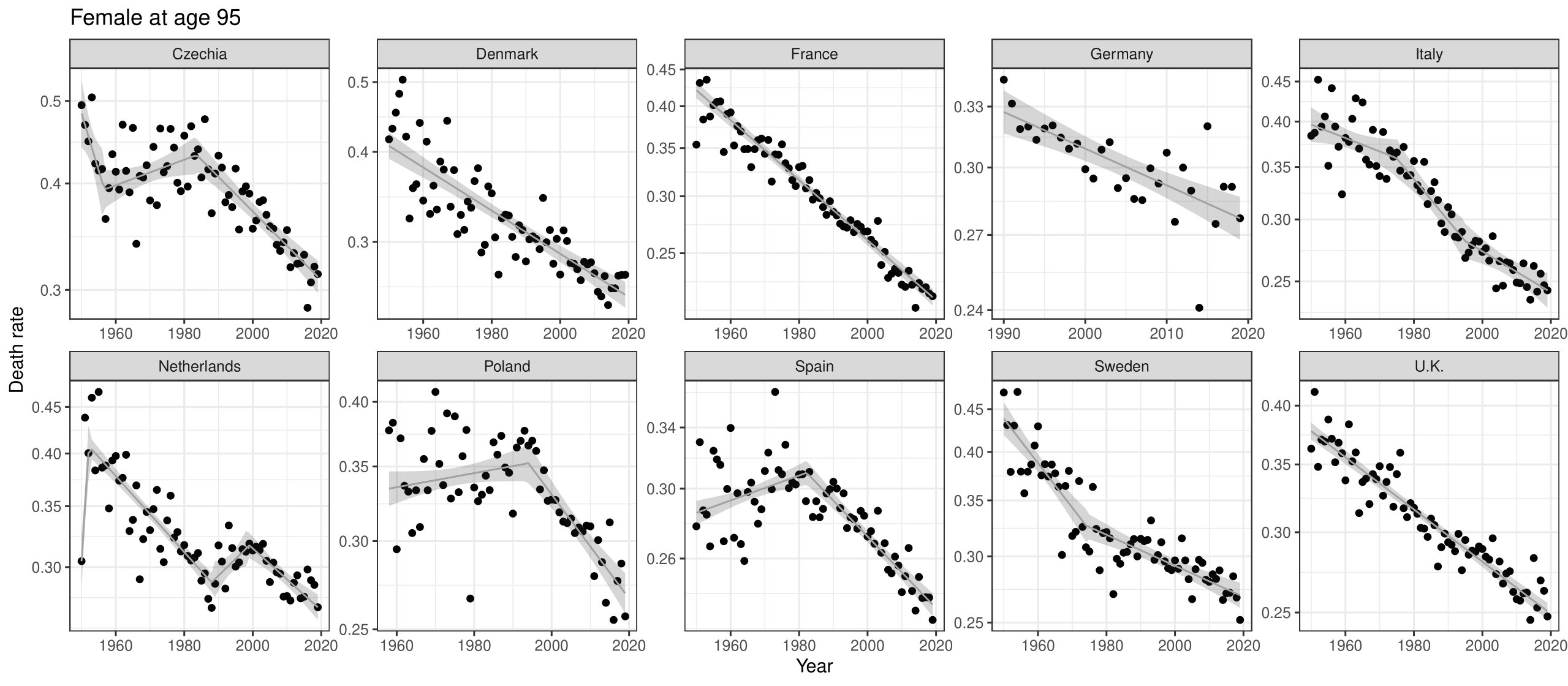}
  \caption{The 1950--2019 (for Germany, 1990--2019) time series of female death rates at age 95 (points), obtained by smoothing the death counts at ages 85 and above by the best-fitting procedure (see Table \ref{rmse}) described in Section \ref{methods}. The solid line represents the best-fitting linear or piecewise linear approximation with its corresponding uncertainty bounds. The existence and number of breakpoints were determined by fitting a segmented quantile regression and testing it (via a likelihood-ratio test) against a linear regression.} \label{f95meanbayes}
\end{sidewaysfigure}

\begin{sidewaysfigure}
  \includegraphics[width=\textwidth]{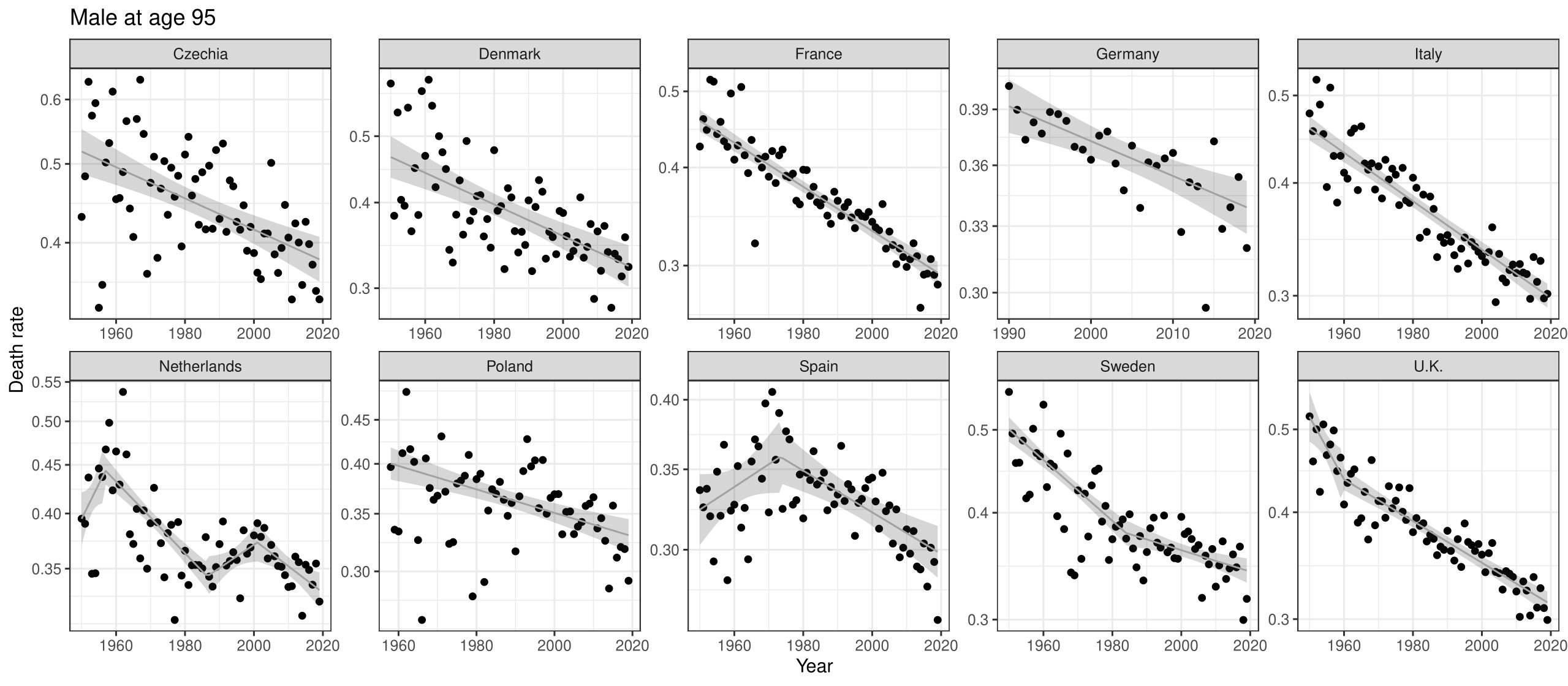}
  \caption{The 1950--2019 (for Germany, 1990--2019) time series of male death rates at age 95 (points), obtained by smoothing the death counts at ages 85 and above by the best-fitting procedure (see Table \ref{rmse}) described in Section \ref{methods}. The solid line represents the best-fitting linear or piecewise linear approximation with its corresponding uncertainty bounds. The existence and number of breakpoints were determined by fitting a segmented quantile regression and testing it (via a likelihood-ratio test) against a linear regression.} \label{m95meanbayes}
\end{sidewaysfigure}

\begin{sidewaysfigure}
  \includegraphics[width=\textwidth]{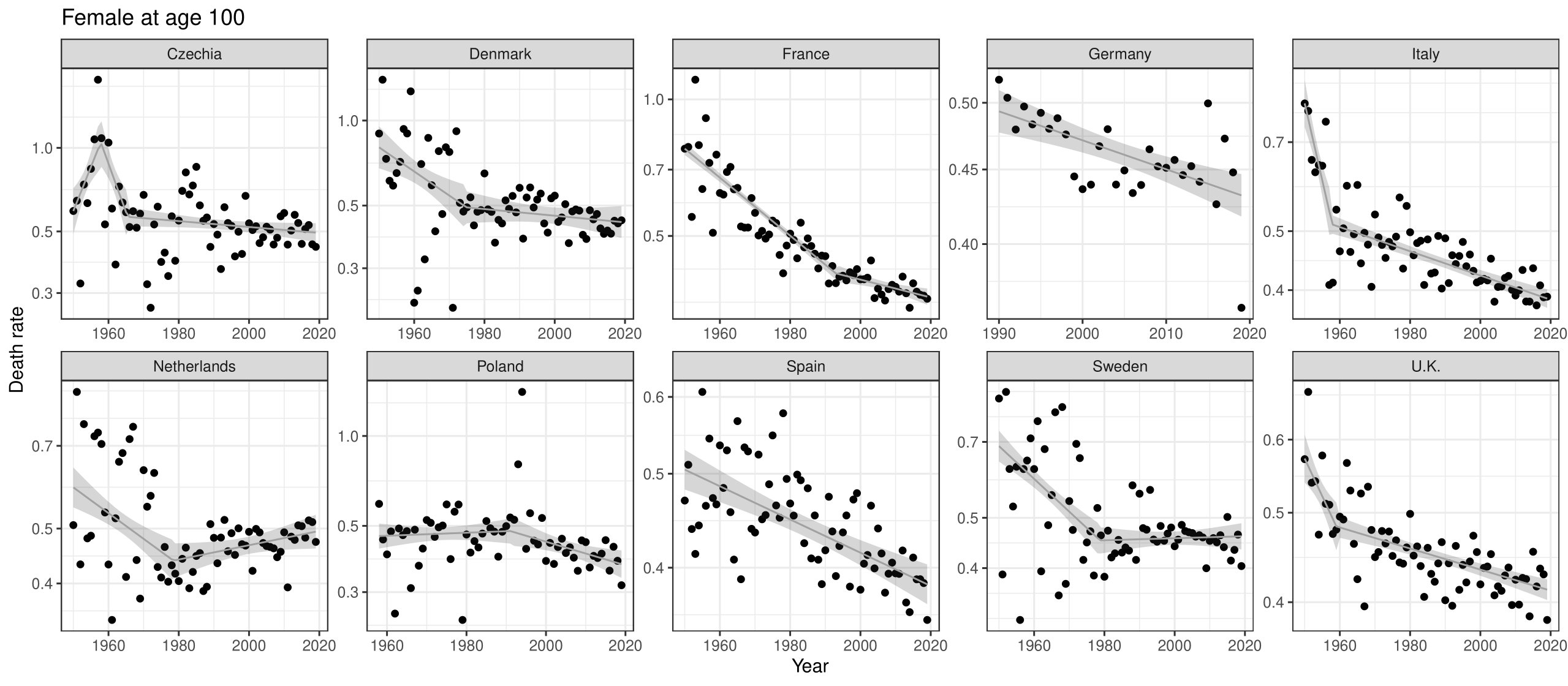}
  \caption{The 1950--2019 (for Germany, 1990--2019) time series of female death rates at age 100 (points), obtained by smoothing the death counts at ages 85 and above by the best-fitting procedure (see Table \ref{rmse}) described in Section \ref{methods}. The solid line represents the best-fitting linear or piecewise linear approximation with its corresponding uncertainty bounds. The existence and number of breakpoints were determined by fitting a segmented quantile regression and testing it (via a likelihood-ratio test) against a linear regression.} \label{f100meanbayes}
\end{sidewaysfigure}

\begin{sidewaysfigure}
  \includegraphics[width=\textwidth]{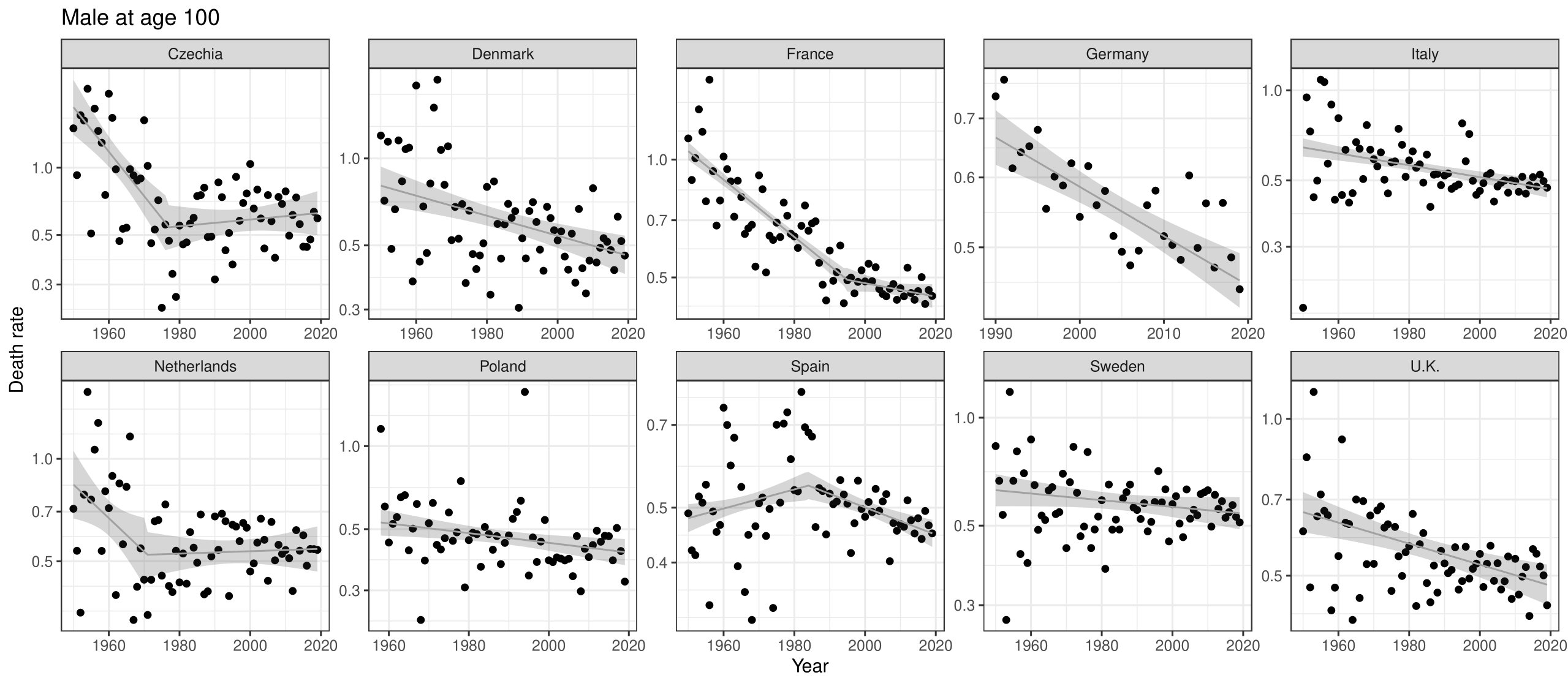}
  \caption{The 1950--2019 (for Germany, 1990--2019) time series of male death rates at age 100 (points), obtained by smoothing the death counts at ages 85 and above by the best-fitting procedure (see Table \ref{rmse}) described in Section \ref{methods}. The solid line represents the best-fitting linear or piecewise linear approximation with its corresponding uncertainty bounds. The existence and number of breakpoints were determined by fitting a segmented quantile regression and testing it (via a likelihood-ratio test) against a linear regression.} \label{m100meanbayes}
\end{sidewaysfigure}

\begin{sidewaysfigure}
  \includegraphics[width=\textwidth]{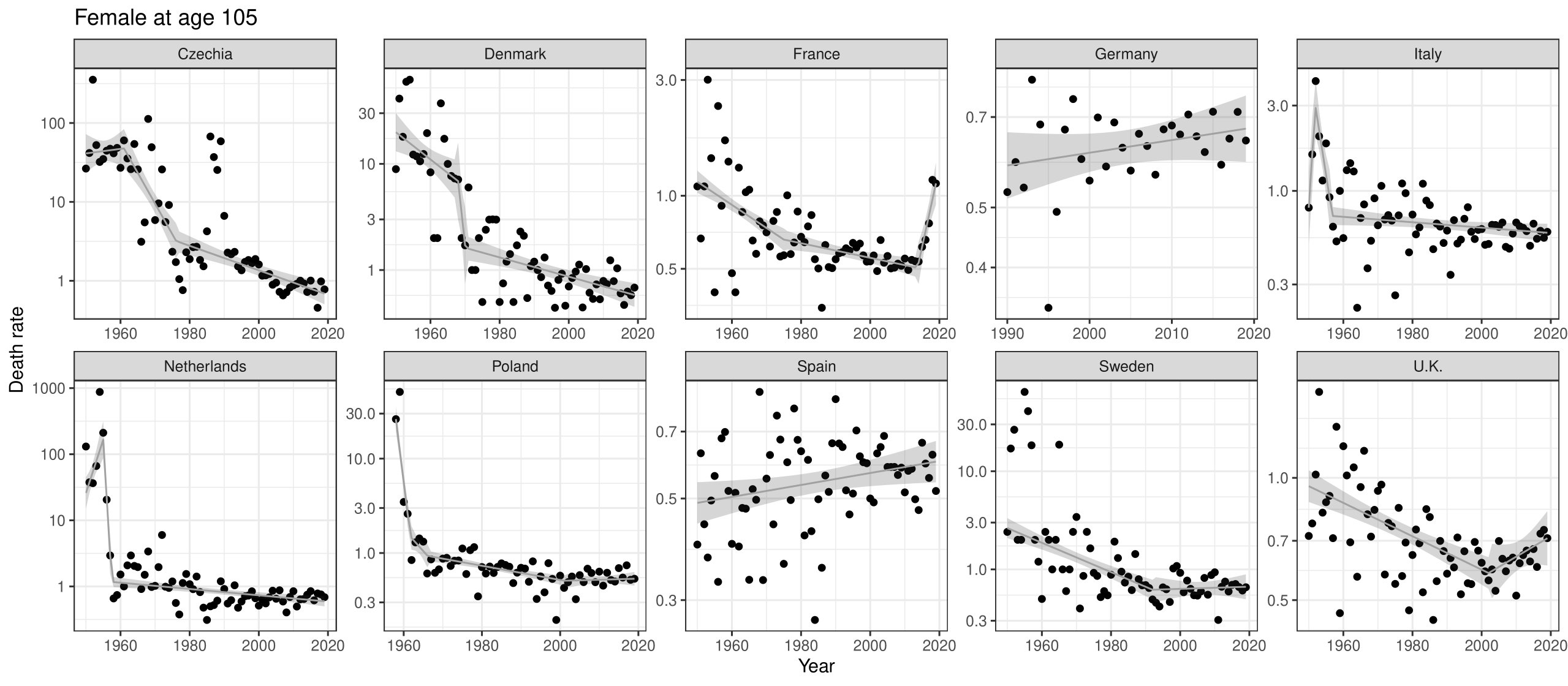}
  \caption{The 1950--2019 (for Germany, 1990--2019) time series of female death rates at age 105 (points), obtained by smoothing the death counts at ages 85 and above by the best-fitting procedure (see Table \ref{rmse}) described in Section \ref{methods}. The solid line represents the best-fitting linear or piecewise linear approximation with its corresponding uncertainty bounds. The existence and number of breakpoints were determined by fitting a segmented quantile regression and testing it (via a likelihood-ratio test) against a linear regression.} \label{f105meanbayes}
\end{sidewaysfigure}

\begin{sidewaysfigure}
  \includegraphics[width=\textwidth]{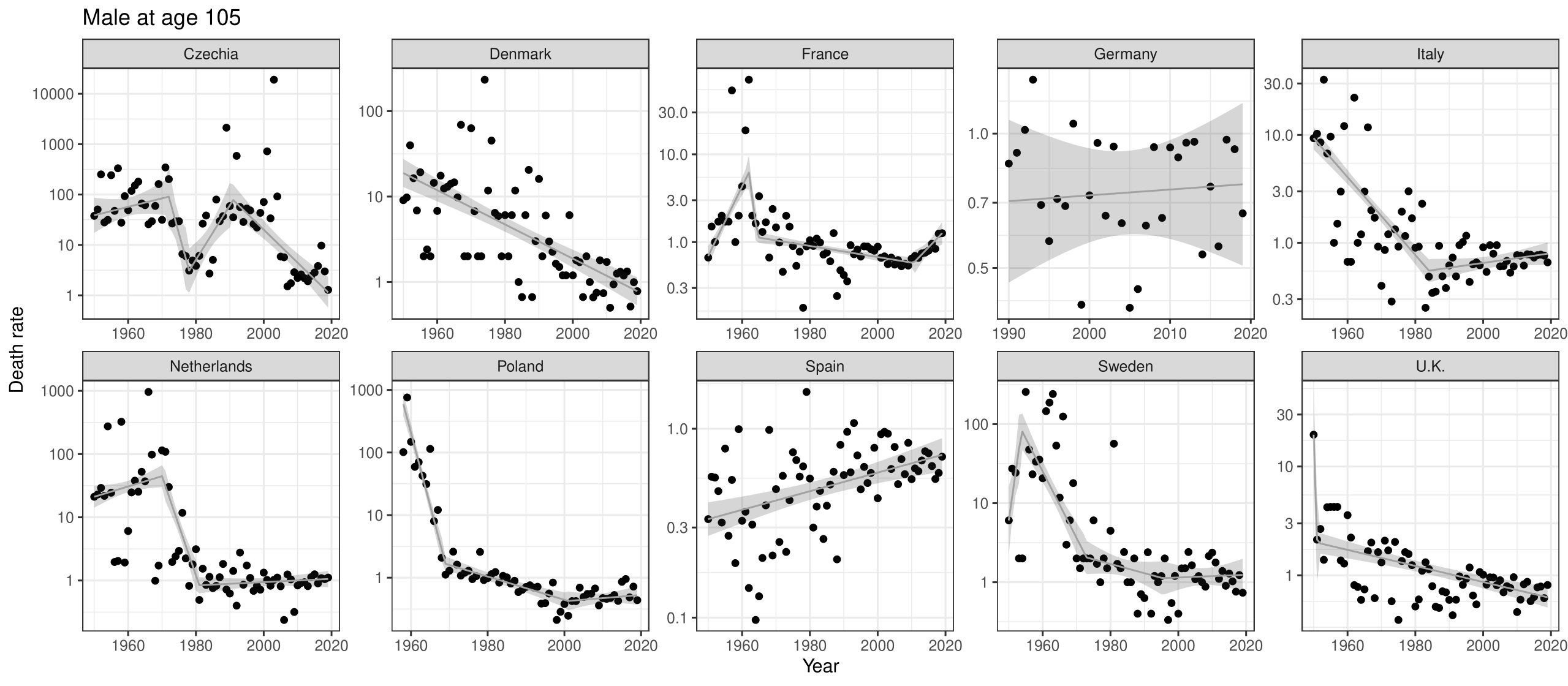}
  \caption{The 1950--2019 (for Germany, 1990--2019) time series of male death rates at age 105 (points), obtained by smoothing the death counts at ages 85 and above by the best-fitting procedure (see Table \ref{rmse}) described in Section \ref{methods}. The solid line represents the best-fitting linear or piecewise linear approximation with its corresponding uncertainty bounds. The existence and number of breakpoints were determined by fitting a segmented quantile regression and testing it (via a likelihood-ratio test) against a linear regression.} \label{m105meanbayes}
\end{sidewaysfigure}

%
%
%

\end{document}